# Why do People Share Misinformation during the COVID-19 Pandemic?

Samuli Laato[1], A.K.M. Najmul Islam[1], Muhammad Nazrul Islam[2] and Eoin Whelan[3]

[1]Department of Future Technologies
University of Turku
Turku, Finland
Correspondence: Samuli Laato sadala@utu.fi

[2]Department of Computer Science and Engineering
Military Institute of Science and Technology
Mirpur Cantonment, Dhaka, Bangladesh
E-mail: nazrul@cse.mist.ac.bd

[3]Business Information Systems
J.E. Cairnes School of Business and Economics
NUI Galway
Ireland

**Abstract:** The World Health Organization have emphasised that misinformation – spreading rapidly through social media – poses a serious threat to the COVID-19 response. Drawing from theories of health perception and cognitive load, we develop and test a research model hypothesizing why people share unverified COVID-19 information through social media. Our findings suggest a person's trust in online information and perceived information overload are strong predictors of unverified information sharing. Furthermore, these factors, along with a person's perceived COVID-19 severity and vulnerability influence cyberchondria. Females were significantly more likely to suffer from cyberchondria, however, males were more likely to share news without fact checking their source. Our findings suggest that to mitigate the spread of COVID-19 misinformation and cyberchondria, measures should be taken to enhance a healthy skepticism of health news while simultaneously guarding against information overload.

# 1. Introduction

*“We’re not just fighting an epidemic; we’re fighting an infodemic”* - WHO Director-General Tedros Adhanom Ghebreyesus[1].

Defined as *“False or inaccurate information, especially that which is deliberately intended to deceive”* (Lazer et al. 2016), misinformation poses a serious threat to public health during pandemics such as the COVID-19 (Zarocostas, 2020). The rapid spreading of such misinformation largely occurs through social media and could result in the lack of adherence to recommended public health measures, or engagement in non-recommended behaviors. One clear example of disseminated misinformation during the COVID-19 suggested 5G cellular network towers contribute to the spread of the virus, reportedly causing people to attack network towers[2]. With several similar cases of misinformation and fake news circulating and being shared on social media, tackling the COVID-19 ‘infodemic’ has been established as a research priority in the WHO Response Strategy[3].

Accurate knowledge and information is essential for human beings to be able to make rational decisions and operate in society (Vosoughi et al., 2018). This is even further emphasized during massively disruptive events, such as the 2020 COVID-19 global pandemic (Bai et al., 2020). The consequences of misinformation spread include fueling cyberchondria (Lewis, 2006), poor health related decisions (Allcott and Gentzkow, 2017) and in the worst cases, deaths due to individuals’ and health official’s inability to accurately evaluate the severity of the situation and take necessary actions (Kata, 2010; Sommerlad, 2020). As evidence of the magnitude of the problem, in 2018 the World Economic Forum listed online misinformation as one of the top 10 global threats to humanity (Howell, 2013; Del Vicario, 2016).

Since the beginning of this century, the quantity and dissemination of misinformation have grown exponentially (Kim and Dennis, 2019). A variety of reasons for have been offered to explain the increased misinformation dissemination, including a decline in social capital, economic inequality, stratification of people into social sub-groups, a distorted and shattered media landscape, and a decline in public trust towards science (Lewandowsky et al., 2017). Typically, the public use of social media is attributed for the rise misinformation (Allcott and Gentzkow, 2017), while websites and traditional news coverage can also contribute to the problem. Fake news is a similar to misinformation, which is defined to be a mimicry of media outlets that lacks the editorial process, norms and journalistic rigor that ensures a trustworthy quality of news reporting (Lazer et al., 2018). The reported increase in misinformation and fake news ultimately led to the term “post-truth” skyrocket in use during 2015-2016 in so that it was selected the Oxford Dictionary’s word of the year in 2016 (Levitin, 2017).

---

1 WHO Director-General Tedros Adhanom Ghebreyesus at the Munich Security Conference on Feb 15.

2 For more information on COVID-19 fake news, see World Health Organization Busting COVID-19 myths at: https://www.who.int/emergencies/diseases/novel-coronavirus-2019/advice-for-public/myth-busters

3 WHO response strategy: http://origin.who.int/blueprint/priority-diseases/key-action/Coronavirus_Roadmap_V9.pdf

The importance of clarity and reliability of information gets highlighted during times where unprecedented events take place and inaccurate and poor decisions have serious consequences (Zarocostas, 2020). The COVID-19 pandemic is a prime example of such an event. The pandemic emerged in late 2019 in Wuhan, China, and quickly spread globally (Chinazzi, et al., 2020) which led to the World Health Organization declaring COVID-19 a global pandemic on March 11th, 2020. As a response to the pandemic, countries issued travel restrictions, closed schools, universities and public services and placed people in quarantine to mitigate the spread of the disease (Fang et al., 2020; Wilder-Smith and Freedman, 2020). While these actions took place, individuals were informed about the pandemic and what individual measures could and should be taken to combat the spread of COVID-19. Social media was filled with sharing COVID-19 related news and articles, some of which were found to be misinformation (Cinelli et al., 2020). The abundance of available information and its ambiguity combined with the novelty of the pandemic situation increased the risk of health anxiety (Doherty-Torstrick et al., 2016; Mathes et al., 2018; White and Horvitz, 2009). Thus, our study incorporates the concept of cyberchondria, which encapsulates the unfounded escalation of concerns about common symptomology based on review of online content. Indeed, cyberchondria may well be the most pressing health issue arising from the abundance of health-related fake news.

To address the COVID-19 misinformation problem, we first need to understand why people share unverified COVID-19 related information through social media. Therefore, from a health perception and information load perspective, the aim of this study is to empirically determine the specific individual drivers of COVID-19 social media misinformation sharing and cyberchondria. Previous studies on why people share fake news or misinformation on social media have found several explanators, of which high trust in online sources is pertinent (Talwar et al., 2019; Khan and Idris, 2019). However, studies about misinformation sharing during a major health-crisis such as COVID-19 are missing. People make different decisions about information when driven by fear and anxiety (Allen et al. 2014). As an example, there is evidence that information-seeking during pandemics is more common in those experiencing worry or fear (Lin et al. 2014). Similarly, previous research suggests that when in a state of fear or distress, peoples' usage patterns and perceptions of social media alters significantly to the extent they can become overloaded and fatigued (Maier et al., 2014;Whelan et al., 2020b). Building on this existing work, we investigate the drivers of COVID-19 specific misinformation and cyberchondria.

To address the research problem, we draw from theories of health perception (i.e. perceived susceptibility and perceived severity) and information load (i.e., information trust and information overload) to develop and test a research model hypothesizing the drivers of COVID-19 misinformation and cyberchondria. We test the model with survey data from 294 Facebook users from Bangladesh using the PLS-SEM analysis technique. The results revealed both information factors are associated with increased cyberchondria and sharing unverified information. The health factors had no impact on sharing unverified information, but did predict increased cyberchondria. Finally, we found no direct relationship between suffering from cyberchondria and sharing unverified information.

# 2. Background

## 2.1 Social Media and Misinformation Sharing

To assess the existing literature pertaining to misinformation and social media, we conducted a network visualization of the relevant papers available through the Scopus database over the past 10 years. The resulting visualization (see Figure 1) shows the literature on misinformation and social media to be grouped into three broad themes. One stream (top left) broadly focuses on applying computer science approaches to detect and prevent the spread of misinformation. A second stream (bottom left) considers the impact of social media misinformation on politics, science, and society in general. The third stream (right) focuses on the relationship between misinformation and population health. It is largely the latter stream from which this study draws from and contributes to.

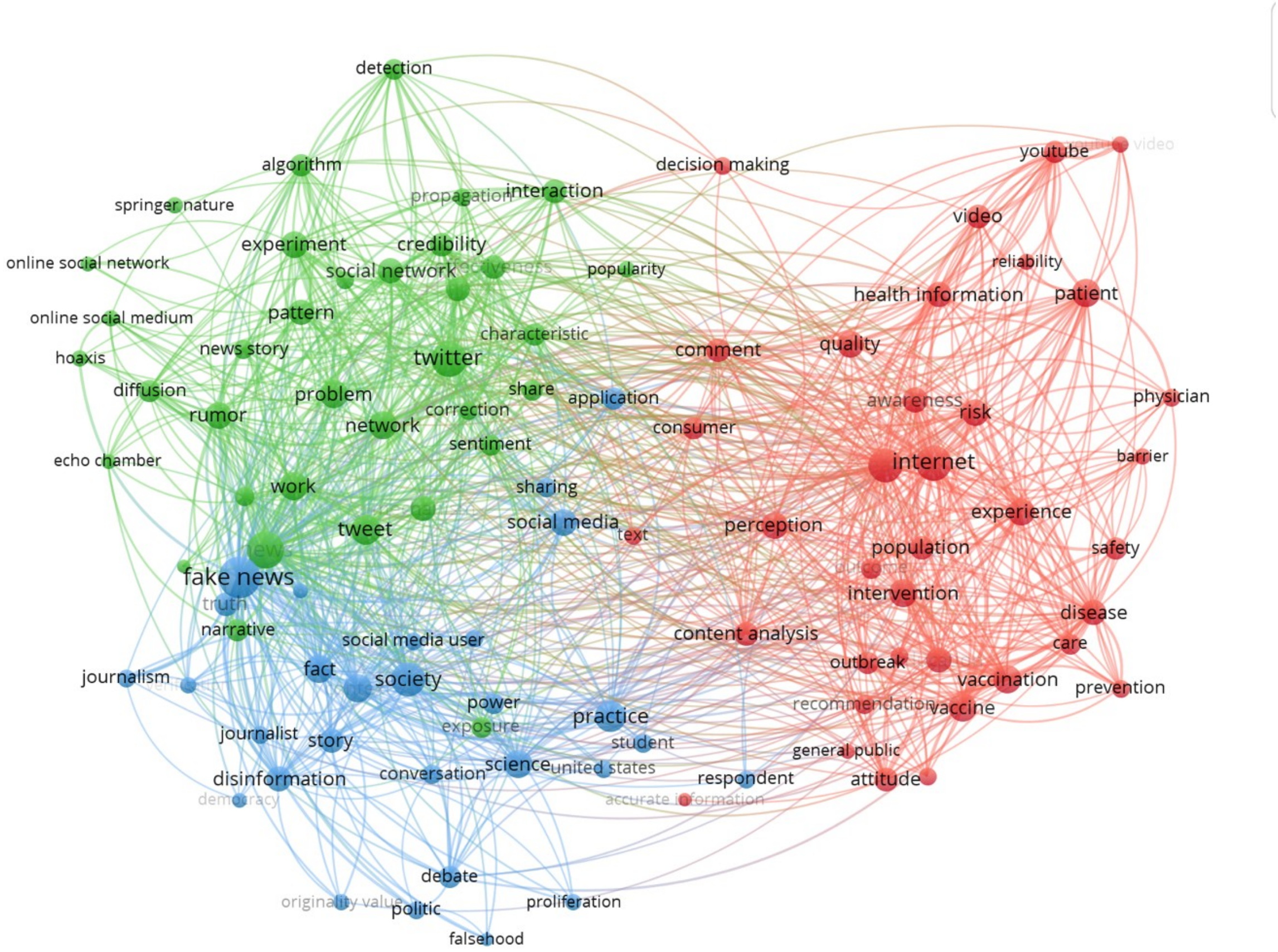

Figure 1. Visualizing the Social Media and Misinformation Literature

While misinformation and fake news can exist anywhere and be mediated through the same channels as any other news, social media in particular has been found to accelerate the spread of fake news (Alcott and Gentzkow, 2016; Shu et al., 2017). Besides sharing news, social media is valuable for people as a way to stay in touch with others (Sponcil and Gitimu, 2013), however, it is also characterized by the desire for self-promotion (Mäntymäki and Islam, 2016). Platforms such as Facebook and Twitter have been blamed for causing polarization with personalized news feeds, but in fact, recent studies have found that human's own choices limit the type of clicked content more than algorithms (Bakashy et al., 2015; Sphor, 2017).

Previous studies have found several mechanisms that affect the spread of fake news on social media, one of which is the use of bot armies to manipulate the platform algorithms to boost the visibility of fake news articles (Lazer et al., 2018; Weedon et al., 2017). Another mechanism relates to people themselves, who can be driven by wishes to either inform or hurt others (Chadwick and Vaccari, 2019). Interestingly, medical professionals themselves were more likely to spread dread rumors than wish ones (Chua and Banerjee 2018). Reports also suggest that some groups of people believe and spread false news, despite better evidence due to ideological reasons (Wolfe, 2002). Studies have argued that the polarizing impact of social media contributes to the spread of fake news via confirmation biases and social influence (Sphor, 2017).

With regards to the number of people circulating fake news articles, almost half of those sharing news articles report to have at some point shared misinformation (Chadwick and Vaccari, 2019). Whether people share the (fake) news articles onward in social media is determined by the relevance, shock value and believability of the content rather than its source (Chadwick and Vaccari, 2019; Chen et al., 2015; Huang et al., 2015). Also lack of experience about online environments and resulting trust in online information, as well as laziness in verifying the information source and lack of skills to do it, are reasons contributing to people sharing misinformation (Khan and Idris, 2019; Talwar et al., 2019). It is almost impossible to accurately determine whether a piece of news is misinformation or based on evidence simply based on the news article itself (Del Vicario et al., 2016), and therefore, additional sources for verifying the news' reliability are needed. Recent studies have demonstrated that by directing readers to pay attention to the news source and its reliability negatively impacts sharing misinformation onwards (Kim and Dennis, 2019).

While quite an amount of empirical investigations have been conducted into the spread of misinformation through social media, there is a dearth of studies, which specifically consider the phenomenon whilst people are in the midst of a pandemic. As previously noted, peoples' decision-making processes are significantly altered when driven by fear and anxiety. Thus, our study will address this gap in our knowledge by specifically focusing on the misinformation drivers within the COVID-19 context.

## 2.2 Cyberchondria

The term cyberchondria is derived from the term, hypochondriasis, which is a condition about excessively and chronically worrying about being seriously ill (Starcevic and Berle, 2015). Hypochondriasis was mixed together with cyber to reflect the cause of this mental state being in the cyberworld, more specifically the internet (Starcevic and Berle, 2013). Thus, cyberchondria is defined as constant online searching for health information which is fuelled by an underlying worry about health that results in increased anxiety (Starcevic and Berle, 2013). Increased time spent searching online for symptoms has been associated with functional impairment and increased anxiety (Doherty-Torstrick et al., 2016; Mathes et al., 2018). Thus, it is clear cyberchondria can be impairing and harmful for individuals (Mathes et al., 2018).

Literature on what causes cyberchondria shows that it is strongly correlated with anxiety. For example, researchers reported that anxiety sensitivity increases cyberchondria (Doherty-Torstrick et al., 2016; Norr et al., 2015). Information overload has been found to be linked to cyberchondria (White and Horvitz, 2009) through the continued seeking of reinforcing information (Norr et al., 2015). Cyberchondria has not been found to be connected to age, gender or even the actual medical status (Fergus and Spada, 2017). However, metacognitive beliefs (Fergus and Spada, 2017) as well as factors such as distaste for ambiguity (McMullan et al., 2019) and intolerance of uncertainty (Norr et al., 2015) also play roles in developing cyberchondria.

As demonstrated by the literature summary on fake news studies on social media, as well as on cyberchondria, previous works have not looked at the relationships between sharing unverified information and cyberchondria. The context of COVID-19 offered us an opportunity for investigating both of these together, as the pandemic escalated into a global health crisis with updates spreading rapidly through social media. Indeed, the lock-down enacted in many countries with workplaces and social activities required to close, may have the unintended consequence of escalating misinformation and cyberchondria as people have more time at their disposal to overload on social media content.

## 2.3 Theoretical Foundation

Understanding cyberchondria and misinformation sharing needs to take into account both health risk and technological factors. Accordingly, for tackling the problem from a wider perspective, we look at three relevant theories, two which incorporate the health behavior aspect: (1) health belief model (HBM); and (2) protection-motivation theory (PMT); and a third which encompasses the impact of technology: (3) cognitive load theory (CLT).

The HBM has been widely used in studies about designing and investigating health behavior change interventions (Eldredge et al., 2016; Orji et al., 2012), but also in other fields such as cybersecurity (Ng et al., 2009). Another theory, which is often used in a similar way as HBM to understand health behavior, is the PMT (Prentice-Dunn and Rogers, 1986). The PMT focuses on understanding the reasons why humans adapt protective health measures, which in the case of the COVID-19 pandemic are, for example, washing hands and self-isolation. However, just like the HBM, the PMT has also been used elsewhere, such as, for understanding why people adopt protective cybersecurity measures (Meso et al., 2013). A review study on PMT identified perceived threat to be the main driver behind protection motivation (Bish and Michie, 2010). For the current study, we employ the concepts of perceived susceptibility and perceived severity to conceptualize perceived threats (i.e. health beliefs) because of their relevance in both HBM and PMT, and because previous studies found these constructs to be the significant predictors for health motivation (Bish and Michie, 2010). Perceived severity is defined as the individual’s appraisal of the severity of the situation with regards to health consequences (Ling et al., 2019) whereas perceived susceptibility is an appraisal of the probability of being vulnerable in the given situation (Ling et al., 2019).

In addition to the theories explaining health behavior, theories accounting for the impact of technology are also required. As we are investigating social media use and information sharing, a theory of particular relevance is the CLT. The CLT is built on the notion that human memory may be divided into biologically primary and secondary knowledge and has limited processing capability (Sweller, 2011). Only small amounts of new information can be processed at a time, with findings suggesting that unclearly structured or too large packets of information make learning and acquiring the knowledge difficult for humans (Sweller, 2011; Paas et al., 2003).

Cognitive overload has been shown to decrease social trust between people (Samson and Kostyszyn, 2015), but also the trust towards AI systems (Zhou et al., 2017). Humans overloaded by information are likely to make careless decisions as they are unable to process surrounding information and experience less self-control (Samson and Kostyszyn, 2015). This was verified in the experiment by Zhou et al., (2017) who found uncertainty presentation to lead to increased trust under small cognitive load, but the presentation of uncertainty under high cognitive load led to decreased trust. Stemming from our review of the CLT literature, the constructs of information trust (Talwar et al., 2019) and information overload (Whelan et al., 2020a) are likely to be salient in explaining misinformation decisions. Hence, these two constructs are central components of our research model.

# 3. Research Model and Hypotheses

## 3.1 Effects of Online Information

The human trust in journalistic information has declined during the past few decades (Lewandowsky et al., 2017). Among theorized causes for this are the internet and social media, which allow people a more direct access to information than what was previously possible (Lewandowsky et al., 2017; Settle, 2018). Through social media, individuals have the potential to detect biases in traditional news reporting but at the same time, are exposed to non-rigorous journalism. Furthermore, it has been documented that algorithms filter only preferred news to individuals, which reinforces existing biases they may have (Bakashy et al., 2015; Sphor, 2017). While this may have increased harmony within social sub-groups, it has simultaneously served to increase inter-group conflict and made people less prepared to hear opposing views (Settle, 2018).

In recent years, a significantly large quantity of fake news and misinformation have been shared on social media (Chadwick and Vaccari, 2019), at times even more frequently than news backed up with journalistic ethics and rigor (Howard et al., 2017). Fake news articles that manage to spread far typically resemble real news to such an extent that it is difficult for both humans and algorithms to distinguish the two from each other (Del Vicario et al., 2016). People who have high trust on online information are increasingly likely to share onward not only real news, but also fake news reports and misinformation (Khan and Idris, 2019, Talwar et al., 2019). Accordingly, we hypothesize the following.

**H1:** Online information trust increases the sharing of unverified COVID-19 information.

Huang et al. (2015) interviewed social media users during the Boston Marathon Bombings and found that the abundance of information and speed of newly occurring events reduced people's ability to verify the information sources. This contributed to an increased spread of misinformation. The finding can be understood through the CLT, which postulates that humans have limited working memory. In novel situations where new information is being presented at high volumes, the human cognitive capacity gets overloaded, which may lead to social media fatigue (Maier et al., 2014; Whelan et al. 2020b) and trigger the evolutionary instinct to retreat to a safer ground, away from the difficult-to-conceptualize information (Sweller, 2011). Once humans are fatigued, it reduces their ability to make sense of the new situation, and hinders their judgement and decision making, for example, with regards to what news is backed up by journalistic rigor and what is not. This implies that when humans are overloaded with information, they are less likely to go through the extra trouble of verifying information sources (Whelan et al., 2020b). Thus, we hypothesize the following.

**H2:** Information overload increases the sharing of unverified COVID-19 information.

In today's world, one should clearly not trust all information they are exposed to on social media. When online information is not critically assessed, cognitive dissonances, cognitive overload, and anxiety emerge (Khan and Idris, 2019; Metzger and Flanagin, 2013; Samson and Kostyszyn, 2015). Despite these apparent negative attributes attached to online information trust, our society and all individuals are dependent on online information (Metzger and Flanagin, 2013). Simply distrusting all online information is not an answer, and instead, cognitive skills on evaluating information sources are needed (Auberry, 2018; Chadwick and Vaccari, 2019). It seems feasible that without the process of cognitive evaluation of the reliability of the online information, combined with the prevalence of online misinformation, trusting online information sources can lead to confirmation bias and give birth to unfounded worries about personal health. Thus, we propose the following.

**H3:** Online information trust increases cyberchondria.

In the case of COVID-19, several factors could contribute to increased cognitive load. First, the situation is new, which forces people to acquire new knowledge. Second, the situation developed fast, forcing humans to adapt to the new knowledge quickly. Third, through social media, individuals across the globe shared their experiences, with lots of news appearing all the time, some real, some fake. The quantity of information further made it difficult to understand the actual state of the situation. Fourth and finally, as the knowledge was being generated and shared rapidly, not all of it could be clearly structured and presented in an optimal and understandable way. The resulting lack of clarity further contributed to cognitive overload. Previous studies of cyberchondria suggest it is associated with information overload (White and Horvitz, 2009) and uncertainty (Norr et al., 2015). Likewise, one' anxiety about their health interacts with the amount of information they seek online to shape health related decisions (Eastin and Gunisler, 2006). Thus, we hypothesize the following;

**H4:** Information overload increases cyberchondria.

## 3.2 Effects of Health Beliefs

The HBM postulates that both perceived susceptibility and perceived severity influence human behavioral responses in the face of health risks (Sheeran and Abraham, 1996). Previous research has not considered how these responses materialize in online and social media behavior. Related research suggests susceptibility and severity will influence the sharing of misinformation through social media. For example, medical professionals are more likely to spread rumors online when the perceived relevance of the rumor to them is high (Chua and Banjeree 2018). When investigating social media use after the Boston Marathon Bombings, Huang et al., (2015) found emotional and physical proximity to increase the likelihood of sharing unverified information. The root cause behind this may be that when people are physically and emotionally closer to the affected area, they would feel a higher level of severity and susceptibility. Therefore, we propose the following two hypotheses.

**H5:** Perceived severity increases unverified COVID-19 information sharing.
**H6:** Perceived susceptibility increases unverified COVID-19 information sharing.

Previous studies on cyberchondria have linked it to health anxiety (Starcevic and Berle, 2013; White and Horvitz, 2009). Cyberchondria differs from health anxiety in that it impairs functionality, but the two are otherwise linked with measurable symptoms (Mathes et al., 2018). Since COVID-19 was declared a global pandemic, most news feeds globally were filled with information about it. The swarm of information released about COVID-19 communicated the severity of the situation, with individuals appraising this information by evaluating the threat as well as their ability to cope with it (Rogers and Prentice-Dunn, 1997). As postulated by the PMT, a natural consequence of a severe threat appraisal is to search for more information on the matter in order to cope with the situation. In practice, this would mean going online to search for more information on COVID-19. The new information could then further accelerate the threat appraisal via the mechanisms already hypothesized earlier. Thus, we propose our two hypotheses.

**H7:** Perceived severity increases cyberchondria.
**H8:** Perceived susceptibility increases cyberchondria.

Finally, we focus on the relationship between cyberchondria and sharing unverified information. From previous studies we know that cyberchondria is a health anxiety issue characterized by repeated and excessive online searches for health information (White and Horvitz, 2009). This exposes to a multitude of online information sources, increasing the likelihood of also encountering fake news and misinformation. Due to the difficulty of distinguishing fake news from real information (Del Vicario et al., 2016), it could also lead to sharing the unverified information. Thus, we propose our final hypothesis.

**H9:** Cyberchondria increases unverified COVID-19 information sharing.

Our final research model connecting the proposed hypotheses is shown in Figure 2.

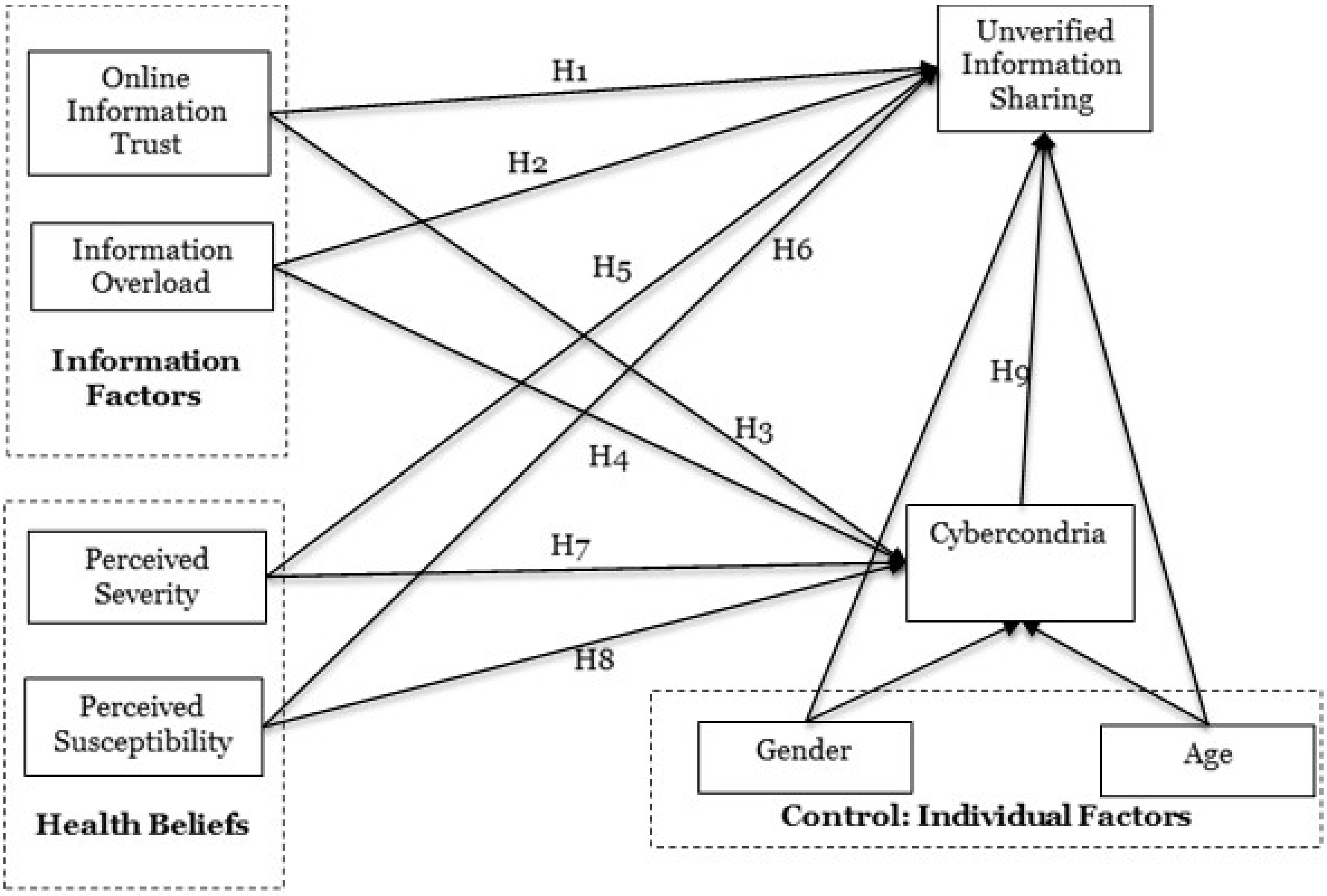

Figure 2. Research Model

# 4. Methodology

## 4.1 Study context

This study was conducted in Bangladesh in March 2020 when the COVID-19 pandemic had been declared a global pandemic by the World Health Organization. To better understand the context, we briefly describe the situation in Bangladesh and the social media use in the country during the data collection period. According to recent statistics by the Internet World Stats, approximately 100 million people in Bangladesh have access to the Internet and around 34 million people use social media[4]. Among the social media users, around 93.28% people use Facebook, 3.31% use Youtube and the remaining people social media platforms are, for example, Twitter, Instagram, Linkedin and Pinterest[5].

## 4.2 Data collection

Data was collected from Bangladeshi social media users via an online survey in March 2020. Most constructs and corresponding survey items in the study were taken from validated scales adapted from prior literature with minor changes to fit with the context. The only exception was unverified COVID-19 information sharing, which was developed for this study.

4 Internet World Stats usage and population statistics, https://www.internetworldstats.com/stats3.htm#asia, (accessed on 8th April, 2020).
5 Social media stats Bangladesh, https://gs.statcounter.com/social-media-stats/all/bangladesh, (accessed on 8th April, 2020).

The survey was distributed to students and faculty of a university in Bangladesh via email, and was available to participate from March 20th until March 31st 2020. We received 299 completed responses, of which 294 were acceptable. Approximately 60% of our respondents were male. All our respondents had accounts in one or more social media platforms. 92% of respondents reported that they use Facebook as one of the main sources to know more about COVID-19, which aligned with the Internet World Stats report about social media use in Bangladesh.

### 4.3 Data analysis and results

We tested the reliability and validity of our data before testing the structural model. We used the PLS-SEM based approach using the tool, SmartPLS for testing reliability and validity as well as testing the structural model. For testing the reliability and validity, we used the thresholds set by Fornell and Larcker (1981).

We then conducted the structural model test. The results are displayed in Figure 3. The model explained 28% variance in cyberchondria and 14% variance in unverified information sharing. Six out of the nine hypothesized relationships were supported by our data. Furthermore, we observed that our control variable gender (1=male, 2=female) had a negative effect on unverified information sharing and a positive effect on cyberchondria. We observed no effect of age on either of our dependent variables.

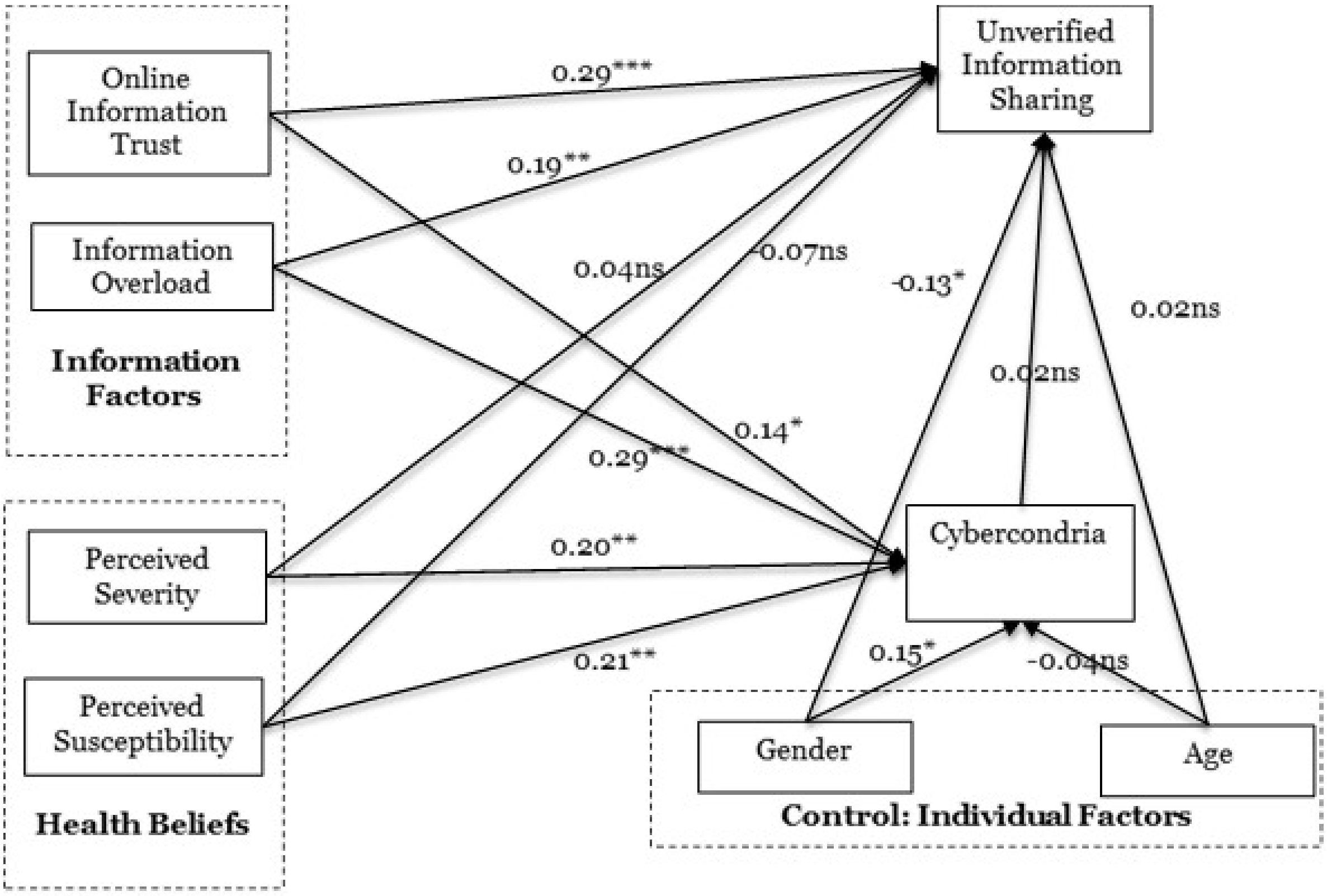

Figure 3. PLS analysis results (***p<0.001; **p<0.01; *p<0.05)

We conducted two post hoc analyses to probe further if gender and age moderates any of our hypothesized relationships. To test the moderating effects, we allowed both age and gender to interact with all the predators of cyberchondria and unverified information sharing. We observed that all interaction terms predicting cyberchondria and unverified information sharing were non-significant except two. The interaction term of information overload and age ($p<0.05$) as well as the interaction term of perceived severity and age ($p<0.05$) had significant negative effects on both cyberchondria and unverified information sharing.

As we observed health belief factors and cyberchondria had non-significant influence on unverified news sharing, we decided to probe it further. Therefore, we conducted another post hoc analysis to investigate the moderating effects of information overload and online trust on these relationships. We observed that information overload actually reinforces the influence of cyberchondria on unverified information sharing ($p<0.05$).

# 5. Discussion

## 5.1 Key Findings

We summarize our key findings as follows.

First, sharing unverified information on COVID-19 was predicted by trust in social media news and social media overload, but surprisingly, not by the measured health threats: perceived severity and perceived susceptibility from COVID-19. The fact that the health beliefs showed no causality to sharing unverified information sharing suggests that while fake news might increase worry about personal health, the worry for personal health does not lead to propagating this news. Our data indicates that people's experience of cyberchondria due to COVID-19 did not influence the sharing of unverified information on social media. However, information overload can reinforce the effect of cyberchondria on unverified information sharing.

Second, we observe that both measured information factors (online information trust and information overload) increased cyberchondria. Information overload had the stronger influence, which seems to suggest that cyberchondria is more fueled by being overwhelmed than trust in online cotent. In addition, both measured health belief factors (perceived severity and perceived susceptibility) increased cyberchondria. Thus, all hypothesizes predicting cyberchondria were confirmed.

Third, we observed that gender had significant effects on both cyberchondria and unverified news sharing. Females experienced higher levels of cyberchondria than males. This finding contrasts with previous research that found gender has no effect on cyberchondria (Fergus and Spada, 2017). The data also suggests that females had a lower tendency to share unverified information on social media compared to their male counterparts. This also contrasts the results of a previous study, which observed females to be more likely to share misinformation (Chen et al., 2015). The effect of age was also measured but it had no significant direct effect on any of the constructs. However, our post hoc analyses showed that age attenuates the effects of information overload and perceived severity on both cyberchondria and unverified information sharing. This suggests that older people experience less cyberchondria and share less unverified information due to experiencing less information overload and perceived severity.

## 5.2 Theoretical implications

Based on our findings we propose three major theoretical implications. First, our work is to the best of our knowledge the first to unite misinformation sharing and cyberchondria together via observing background factors impacting both. In addition, we observed these factors during the COVID-19 global pandemic, which offered a novel research context. Accordingly, this work opens up a new unexplored research area and combines theories from both health behavior literature and instructional science to understand the studied relationships. Previous literature on fake news and misinformation has focused mainly on how to detect fake news using algorithmic means or the motives behind creating and sharing fake news (Shu et al., 2017). Our study offers perspectives into how information overload during novel and unprecedented situations might accelerate the propagation of fake news due to the human factor. Our paper initiates new discussions on identifying, but also controlling the underlying factors that contribute to the spread of fake news during global crises such as the COVID-19 pandemic. Moreover, our study confirms cyberchondria to be a side effect of the COVID-19 pandemic.

Second, developing a new construct is seen as a major contribution in information systems research (Mäntymäki et al., 2020). In this paper, we developed a new construct, namely unverified information sharing, applied to COVID-19, to capture how social media users may propagate fake news or misinformation without authenticating the information. Therefore, we contribute to the literature on fake news (e.g. Del Vicario et al., 2016; Howard et al., 2017) by providing a validated scale.

Third, we identified several novel associations in our study. We found that online information trust and information overload are the two main antecedents of sharing unverified information on social media. Talwar et al. (2019) found online trust as the most important antecedent of fake news sharing. Khan and Idris (2019) also reported possible association between higher levels of trust and unverified information sharing. We confirm the findings of these prior literature in the context of the COVID-19 pandemic. At the same time, we extend the prior literature (Talwar et al., 2019; Khan and Idris, 2019) by showing information overload as another main antecedent of sharing unverified information on social media. Huang et al. (2015) in their interview-based research concluded that information overload was related to fake news sharing during Boston bombings. Our study verifies this finding using a quantitative approach in the context of COVID-19. We also identified four factors, namely online information trust, information overload, perceived severity and perceived susceptibility that had a positive correlation with cyberchondria

## 5.3 Practical implications

Based on our findings, intervention strategies such as nudging people to consider the source of social media content, and to consume manageable amounts of that content, are likely to be effective in reducing the spread of misinformation and cyberchondria in crisis situations. While nudging interventions have been found to be effective when dealing with artificially created and benign misinformation (e.g. celebrity gossip), their efficacy when applied to real and personally involved crises have yet to be empirically tested (Kim and Dennis, 2019).. Additionally, due to COVID-19, many people are out of work or unable to partake in social activities, and thus have more time to consume social media content. Information overload may well be an unintended consequence of the COVID-19 crisis which exacerbates the problems of misinformation and cyberchondria. Health organizations can use our findings to educate social media users to consume content in a sustainable manner and thus avoid these problems. Likewise, social media companies have a significant role to play in curbing COVID-19 misinformation. WhatsApp have already introduced restrictions on the forwarding of messages. Our findings suggest that if social media companies restricted the amount of COVID-19 specific information people are exposed to, this would be effective in curbing misinformation and cyberchondria problems.

## 5.4 Limitations and Future Research

As a cross-sectional survey, our results did not account for any change that might have occurred in the observed behavior during the COVID-19 pandemic. Answers to the survey were collected from university educated persons in Bangladesh who were using social media. As such, the results might not be representative of the entirety of the Bangladeshi or world population. In fact, education level might be one important factor that can reduce the sharing of unverified information, as previous literature suggests increasing the population's education and digital literacy levels specifically, to be the answer to the post-truth era and abundance of fake news (Auberry, 2018; Chadwick and Vaccari, 2019; Ireland, 2018).

Lewandowsky et al. (2017) argue that research investigating misinformation or its impacts should be situated within a wide context, taking into account technological, political and societal factors. Looking at our study from this perspective, we measured the technological factors such as information overload, however, we did not account for political or societal factors. Thus, future research could expand on the current study by taking into account the political and societal dimensions. In practise, this could mean further investigation into the role, responsibility and ability of governments and platform developers to direct social media users towards trustworthy clear information, warding against information overload and consequently cyberchondria, as well as impulses to read and share fake news. On a societal level, future research agenda can include looking at the impact of cyberchondria on individual well-being during global pandemic crises such as the COVID-19, and designing measures for mitigating the negative impacts.

Samson and Kostyszyn (2015) proposed that cognitive overload is one of the causes for the observed increase in mistrust, and that trust can be increased by reducing cognitive load. Cognitive load has also obvious effects on perceptions on information, including health information, and information overload (Sweller, 2011). In the current work we did not measure the respondent's cognitive load during COVID-19 and reading online information, but future work could expand on the model by taking into account the impact of cognitive load on both the health and information factors. We find the CLT promising in explaining our findings and invite practitioners as well as scholar to investigate whether efforts to reduce cognitive load during pandemics can alleviate both the sharing of fake news and cyberchondria.

## References

***Note: All references are not up to date***

# Appendix

Will be provided in the final publication.